# Flexoelectricity induced spatially modulated phases in ferroics and liquid crystals (Author review)

Anna N. Morozovska[1*], Victoria V. Khist[2], Maya D. Glinchuk[3], Christian M. Scherbakov[4], Maxim V. Silibin[5], Dmitry V. Karpinsky[6], and Eugene A. Eliseev[3†]

[1]*Institute of Physics, National Academy of Sciences of Ukraine,*
*46, pr. Nauky, 03028 Kyiv, Ukraine*

[2]*Institute of Magnetism, National Academy of Sciences of Ukraine and Ministry of Education and Science of Ukraine,*

[3] *Institute for Problems of Materials Science, National Academy of Sciences of Ukraine,*
*3, Krjijanovskogo, 03142 Kyiv, Ukraine,*

[4] *Taras Shevchenko Kiev National University, Physical Faculty, Chair of Theoretical Physics,*
*4e, pr. Akademika Hlushkova, 03022 Kyiv, Ukraine*

[5] *National Research University of Electronic Technology "MIET",*
*Moscow, Zelenograd, Russia*

[6] *Scientific-Practical Materials Research Centre of NAS of Belarus, Minsk, Belarus*

**Abstract**

In the review we briefly analyze the state-of-art in the theory of flexoelectric phenomena and analyze how significantly the flexoelectric coupling can change the polar order parameter distribution in different ferroics and liquid crystals. The special attention in paid to the appearance of the spatially modulated phases induced by the flexocoupling in condensed and soft matter. Results of theoretical modeling performed in the framework of the Landau-Ginzburg-Devonshire formalism revealed that the general feature, inherent to both ferroics and liquid crystals, is the appearance of the spatially-modulated phases is taking place with increasing of the flexocoupling strength. We'd like to underline that theoretical and experimental study of flexoelectricity and related phenomena in nanosized and bulk ferroics, liquid crystals and related materials are very important for their advanced applications in nanoelectronics, memory devices and LC displays.

---

[*] corresponding author, e-mail: anna.n.morozovska@gmail.com

[†] corresponding author, e-mail: eugene.a.eliseev@gmail.com

# 1. Flexoelectric effect in ferroics

The flexoelectric effect, first predicted theoretically by Mashkevich and Tolpygo [1] in 1957, exists in any matter (condensed or soft one), making the effect universal [2, 3, 4, 5, 6]. The static ***flexoelectric effect*** is an electric polarization generated in solids by a strain gradient and vice versa, whereas broadly known piezoelectricity assume homogeneous strain conditions. The induced strain is linearly proportional to the polarization gradient and, and the proportionality coefficients *f*, which are the components of the flexocoupling tensor, are fundamentally quite small, $f \sim e/a$, where *e* and *a* are respectively electronic charge and lattice constant [7]. Rigorously, the direct and converse static flexoelectric coupling constants $f_{ijkl}$ were described with a fourth-rank tensor, as [2-6]:

$$P_i^{flexo} = f_{ijkl}\frac{\partial u_{jk}}{\partial x_l} \qquad u_{ij} = f_{ijkl}\frac{\partial P_k^{flexo}}{\partial x_l}. \qquad (1)$$

in these expressions $u_{jk}$ and $P_i^{flexo}$ are the tensor strain and polarization vector components, respectively. The physical picture of the flexoelectric effect in solids is shown in **Figs. 1(a)-(c) (a)** When the geometric centres of positive and negative charges coincide, the net dipole moment of the unit cell is zero, and corresponding unstrained 2D structure of elementary charges is shown in **Fig.1(a).** When each unit cell is uniformly tensiled, but the tension gradually varies from one cell to another. The cations are displaced from the centre of the deformed unit cell the strain-gradient induces an uncompensated dipole moment via the flexocoupling mechanism [**Fig.1(b)**]. An inhomogeneous deformation of the unit cell also produces a net dipole moment via the flexocoupling effect [**Fig.1(c)**].

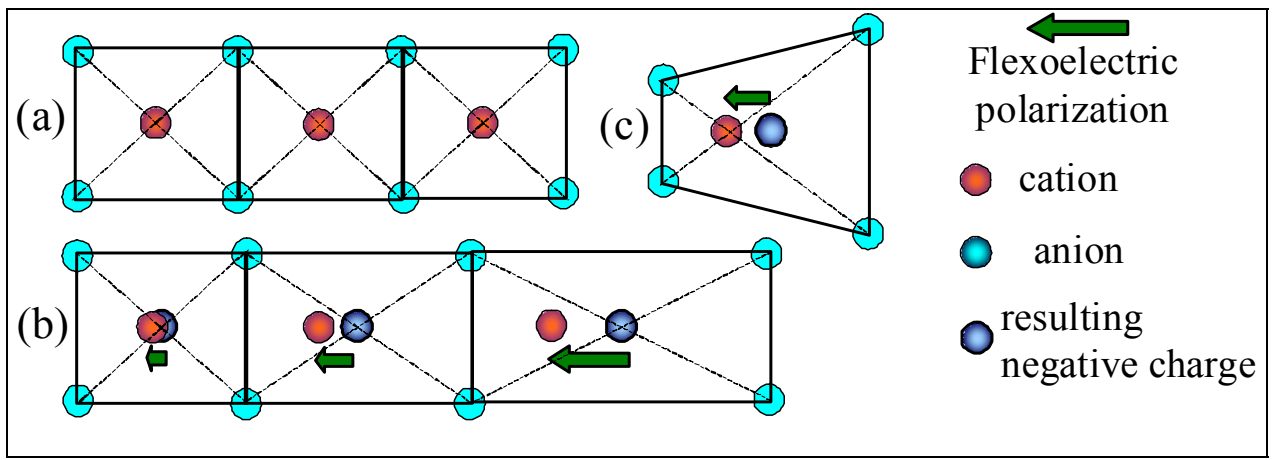

**Figure 1. Flexoelectric effect origin in solids.** (a) Unstrained 2D structure of elementary charges with zero net dipole moment. (b) Each unit cell is uniformly tensiled, but the tension gradually varies from one cell to another. An uncompensated dipole moment appears via the flexocoupling mechanism. (c) An inhomogeneous deformation of the unit cell produces a net dipole moment via the flexocoupling effect. (Adapted from the papers [5, 11-13])

Unlike piezoelectricity, which exists only in noncentrosymmetric systems of 20 point groups, flexoelectricity occurs in all 32 crystalline point groups, because the strain gradients break inversion symmetry. Owing to the universal nature, flexoelectricity permanently attracts broad scientific interest, but its application potential in macro-materials is fundamentally limited due to the small strength $f$.

The flexoelectricity impact is of great importance in nanosized ferroics [8], for which the evolved elastic strain and stress gradients are omnipresent near the surfaces, in thin films [9, 10, 11], at the topological defects, such as domain wall, and near the interfaces [10, 12, 13, 14]. Owing to the surface and interface effects, the flexoelectric effect appears spontaneously near the surface of thin films and nanoparticles [15]; everywhere, where the electric polarization distribution is inhomogeneous. Notably, that the influence of flexoelectricity is important not only in thin ferroic films and nanoparticles but also in their micro- and nanograined ceramics [16, 17]. Despite the great importance of flexoelectricity, its tensorial components strength, estimated by Kogan [7] in early 1963, remained poorly known for most of ferroics up to now [18], except for the experimental measurements [19, 20, 21] and *ab initio* calculations [22, 23] of some components for ferroelectric perovskites, and fundamental limits on the coefficients upper bonds [24].

The direct and converse static flexoelectric effects, which lead to the appearance of polarization due to the strain gradient and vise versa [1], exist in a ferroic of arbitrary symmetry [8, 5, 6], and corresponding tensorial Lifshitz invariant $\frac{f_{klij}}{2}\left(P_i \frac{\partial u_{kl}}{\partial x_j} - u_{kl} \frac{\partial P_i}{\partial x_j}\right)$ should be included to the free energy functional of all those ferroics, for which the polarization component(s) $P_i$ is a primary order parameter. Using the way it was shown that the flexocoupling term in the form of Lifshitz invariant can induce thermodynamically stable incommensurate spatially modulated phase (**SMP**) in the temperature range between the disordered parent phase (**PP**) and long-range ordered homogeneous phase (**HP**) in many ferroics [25, 26, 27, 28]. Note that the static flexoelectric effect is omnipresent from the symmetry theory considerations, and the earliest [1] and recent [22, 23, 29] microscopic calculations give nonzero values of flexoelectric coefficients $f_{klij}$ for many ferroelectrics. Also it becomes possible to define the static flexoelectric coefficients from direct experiments [20, 21, 30], as well as from the fitting of soft phonon spectra in ferroelectrics (see e.g. [31] and refs. therein).

Notably, that chirality, being the property of an object to be incompatible with its mirror image, can be strongly affected by the flexoelectric coupling. In particular, bichiral structure of ferroelectric domain walls can be driven by flexoelectric coupling [32] and chiral-achiral phase transitions at the walls becomes possible [33].

From considerations of the symmetry theory stating that all terms and invariants, which existence does not violate the symmetry of the system, are allowed, Kvasov and Tagantsev et al [34] predicted the existence of a cross-term in the kinetic energy, $M_{ij}\frac{\partial P_i}{\partial t}\frac{\partial U_j}{\partial t}$, and named it dynamic flexoelectric effect (see reviews [3, 4] and refs therein). Here $M_{ij}$ is the strength of dynamic flexoelectric coupling and $U_i$ is elastic displacement. At present the situation with the magnitudes $M_{ij}$ of dynamic flexoeffect is more complex and controversial that for the static one, because there are microscopic theories in which the effect is absent [35]. The Stengel result [35] contradicts to Kvasov and Tagantsev result [34], who evaluated the strength of the dynamic flexoelectric effect in SrTiO$_3$ from microscopic calculations and it appeared comparable to that of the static bulk flexoelectric effect. More discussion of the problem can be found in Refs.[31, 36].

**2. Impact of the flexocoupling on phonon spectra and spatially modulated phases in ferroics**

Investigation of dynamic characteristics of phase transitions in ferroics, such as their soft phonon spectra, attracts great attention of scientists for many years, being the source of valuable information for fundamental physical research and advanced applications [37]. For ferroelectrics the frequency $\omega_{TO}$ of transverse optic (**TO**) soft mode depends on temperature *T*, at that $\omega_{TO}(T_C) = 0$ at transition temperature $T = T_C$ [38].

Basic experimental methods, which contain information about the soft modes and spatial modulation of the order parameter in ferroics (such as antiferroelectrics, proper and incipient ferroelectrics) are dielectric measurements [39], inelastic neutron scattering [38, 40, 41, 42, 43, 44], X-ray [45, 25, 46, 47], Raman [48] and Brillouin [45, 46, 49, 50, 51, 52] scatterings and ultrasonic pulse-echo method [49, 51] allowing hypersound spectroscopic measurements. Scattering experiments proved that not only the TO mode softens substantially with decreasing temperature to freeze out at $T_C$ in ferroics (such as ferroelectric perovskites), but also finite wave vector anomalies appear in the transverse acoustic (**TA**) mode for structural phase transitions [53, 54, 55].

Using the LGD theory [56, 57, 58], Morozovska et al. derived analytical expressions for the soft phonon modes frequency $\omega(k)$ dependence on the wave vector *k* and examined the conditions of the soft acoustic TA-modes appearance in ferroelectrics depending on the magnitude of the flexoelectric coefficient *f* and temperature *T*. If the magnitude of the flexoelectric coefficient *f* is equal to the temperature-dependent critical value $f^{cr}(T)$ at the temperature $T = T_{IC}$, $|f| = f^{cr}(T_{IC})$, then the TA-mode frequency tends to zero at $k \rightarrow k_0^{cr}$ according to the linear law $\omega(k \rightarrow k_0^{cr}) \sim |k - k_0^{cr}|$ and, simultaneously, the ferroelectric polarization becomes spatially modulated. When the magnitude of the

flexocoefficient is more than the critical value $|f| > f^{cr}(T)$ in a temperature range lower than Curie temperature $T_C$, $T_C < T < T_{IC}$, corresponding to the incommensurate phase, the TA-mode becomes zero for two wave vectors $k = k_{1,2}^{cr}$ according to the squire root law, $\omega(k \to k_{1,2}^{cr}) \sim \sqrt{|k_{1,2}^{cr} - k|}$, and does not exist in the range of wave vectors $k_1^{cr} < k < k_2^{cr}$, where the incommensurate modulation exists. At fixed flexocoefficient $f$ the transition into the incommensurate phase can appear at the temperature $T_{IC}$ that depends on $f$. In addition we predicted the appearance of the "rippled" flexocoupling-induced incommensurate phase in the initially commensurate ferroics. The available experimental data on hypersound velocity in the solid solutions $Sn_2P_2(S,Se)_6$ [49] and neutron scattering in organic ferroelectric $(CH_3)_3NCH_2COO \cdot CaCl_2 \cdot 2H_2O$ [43] are in a semi-quantitative agreement with the theoretical results [56-58]. For improvement and for quantification of the theory, it is necessary to measure the frequency dependence of the TA-modes in a uniaxial ferroelectric with a spatially modulated phase in the temperature interval near its appearance.

### 3. General formulation of Landau-Ginzburg-Devonshire formalism

Landau-Ginzburg-Devonshire (**LGD**) expansion of bulk ($F_V$) part of Helmholtz free energy $F$ on the order parameter $\mathbf{\eta}$ and strain tensor components $u_{ij}$ has the form:

$$F_V = \int_V d^3r \begin{pmatrix} \dfrac{a_{ij}(T)}{2}\eta_i\eta_j + \dfrac{a_{ijkl}}{4}\eta_i\eta_j\eta_k\eta_l + \dfrac{a_{ijklmn}}{6}\eta_i\eta_j\eta_k\eta_l\eta_m\eta_n - \eta_i\left(E_{0i} + \dfrac{E_i^d}{2}\right) \\ + \dfrac{g_{ijkl}}{2}\left(\dfrac{\partial \eta_i}{\partial x_j}\dfrac{\partial \eta_k}{\partial x_l}\right) + \dfrac{w_{ijkl}}{2}\left(\dfrac{\partial^2 \eta_i}{\partial x_j^2}\dfrac{\partial^2 \eta_k}{\partial x_l^2}\right) + \dfrac{h_{ijk}}{2}\eta_i^2\left(\dfrac{\partial \eta_j}{\partial x_k}\right)^2 \\ - \dfrac{f_{ijkl}}{2}\left(\eta_k\dfrac{\partial u_{ij}}{\partial x_l} - u_{ij}\dfrac{\partial \eta_k}{\partial x_l}\right) - q_{ijkl}u_{ij}\eta_k\eta_l + \dfrac{c_{ijkl}}{2}u_{ij}u_{kl} + \dfrac{v_{ijklmn}}{2}\left(\dfrac{\partial u_{ij}}{\partial x_m}\dfrac{\partial u_{kl}}{\partial x_n}\right) \end{pmatrix} \quad (2a)$$

Coefficients $a_{ij}(T)$ explicitly depend on temperature $T$. Coefficients $a_{ij}^S$, $a_{ijkl}$, $a_{ijkl}^S$ are supposed to be temperature independent, constants $g_{ijkl}$ and $v_{ijklmn}$ determine magnitude of the gradient energy. Tensors $w_{ijkl}$ and $a_{ijkl}$ are positively defined, $q_{ijkl}$ is the bulk striction coefficients; $c_{ijkl}$ are components of elastic stiffness tensor [59]; $f_{ijkl}$ are the components of flexocoupling tensor. In fact, only the Lifshitz invariant $\dfrac{f_{ijkl}}{2}\left(\eta_k\dfrac{\partial u_{ij}}{\partial x_l} - u_{ij}\dfrac{\partial \eta_k}{\partial x_l}\right)$ is relevant for the bulk contribution due to the surface energy presence in the spatially confined ferroics. The gradient terms like $v_{ijklmn}(\partial u_{ij}/\partial x_k)(\partial u_{lm}/\partial x_n)$ are responsible for the stable smooth distribution of the order parameter at nonzero strain gradients, since the presence of Lifshitz invariant essentially changes the stability conditions [60].

Lagrange function is $L = \int_t dt (F_V - E_V)$, where the free energy $F_V$ is given by Eq.(2a), and kinetic energy $E$ is given by expression

$$E = \int_V d^3r \left( \frac{\mu}{2}\left(\frac{\partial \eta_i}{\partial t}\right)^2 + M_{ij}\frac{\partial \eta_i}{\partial t}\frac{\partial U_j}{\partial t} + \frac{\rho}{2}\left(\frac{\partial U_i}{\partial t}\right)^2 \right), \quad (2b)$$

The kinetic energy includes the dynamic flexocoupling, $M_{ij}\frac{\partial \eta_i}{\partial t}\frac{\partial U_j}{\partial t}$, relating the order parameter $\eta_i$ and elastic displacement $U_i$. Its tensorial strength is $M_{ij}$. $\rho$ is the density of a ferroelectric. The strain is $u_{ij} = \frac{1}{2}\left(\frac{\partial U_i}{\partial x_j} + \frac{\partial U_j}{\partial x_i}\right)$. Dynamic equations of state follows have the form of Euler-Lagrange equations allowing for the possible Khalatnikov-type relaxation of the order parameter:

$$\Gamma \frac{\partial \eta_i}{\partial t} = -\frac{\delta L}{\delta \eta_i}, \qquad \frac{\delta L}{\delta U_i} = 0. \quad (3)$$

### 4. Analytical solutions for commensurate and incommensurate ferroics

Thermodynamic analyses of analytical solutions for commensurate and incommensurate ferroics basing on Eqs.(2) can be simplified keeping in mind that the contribution of the terms, which magnitude is proportional to the diagonal flexocoupling constants $f_{iiii}$ used to be much smaller than those proportional to the nondiagonal ones, $f_{ijij}$ ($i \neq j$), due to the strong depolarization effect for diagonal modes [6, 56]. Keeping in mind that we are looking for large enough flexoeffect contribution ($|f| > f^{cr}(T)$), we are going to consider the nondiagonal terms ($\sim f_{ijij}$) only, and so the Lifshitz invariant asquires the form $\frac{f_{ijij}}{2}\left(\eta_i \frac{\partial u_{ji}}{\partial x_j} - u_{ji}\frac{\partial \eta_i}{\partial x_j}\right)$. It what follows let us denote that $\eta_i = \eta$, $f_{ijij} \equiv f$ and $u_{ij} \equiv u$ [57, 58]. Thus Eq.(2a) can rewritten in the form [57, 58]:

$$F_V = \int_L dx \left( \begin{array}{c} \frac{\alpha(T)}{2}\eta^2 + \frac{\beta}{4}\eta^4 + \frac{\gamma}{4}\eta^6 - \eta E + \frac{g}{2}\left(\frac{\partial \eta}{\partial x}\right)^2 + \frac{w}{2}\left(\frac{\partial^2 \eta}{\partial x^2}\right)^2 + \frac{h}{2}\eta^2\left(\frac{\partial \eta}{\partial x}\right)^2 \\ -\frac{f}{2}\left(\eta\frac{\partial u}{\partial x} - u\frac{\partial \eta}{\partial x}\right) - qu\eta^2 + \frac{c}{2}u^2 + \frac{v}{2}\left(\frac{\partial u}{\partial x}\right)^2 \end{array} \right) \quad (4)$$

Typically the nonlinear gradient parameter $h$ is small and its influence will be neglected hereinafter. Thermodynamic equations of state obtained from the variation of the free energy (4) on order parameter $\eta$ and strain $u$ have the form [57]:

$$\alpha\eta + \beta\eta^3 + \gamma\eta^5 - g\frac{\partial^2 \eta}{\partial x^2} + w\frac{\partial^4 \eta}{\partial x^4} - E - f\frac{\partial u}{\partial x} - 2qu\eta = 0, \quad (5a)$$

$$cu - v\frac{\partial^2 u}{\partial x^2} + f\frac{\partial \eta}{\partial x} - q\eta^2 = 0. \qquad (5b)$$

The solution of these equations can be found after their linearization in the vicinity of spontaneous values $\eta = \eta_S + \int dk\, \exp(ikx)\tilde{\eta}$ and $u = u_S + \int dk\, \exp(ikx)\tilde{u}$ in k-domain. Perturbation field has a conventional form $E = \int dk\, \exp(ikx)\tilde{E}$. In a high temperature parent phase $\alpha > 0$, $\eta_S = 0$, $u_S = 0$. In a low temperature ordered phase, where $\alpha < 0$, $\eta_S \neq 0$, $u_S \neq 0$, the spontaneous strain and order parameter values can be determined from Eqs.(5) at zero gradients, namely $\alpha\eta_S + \beta\eta_S^3 + \gamma\eta_S^5 - 2qu_S\eta_S = 0$ and $cu_S - q\eta_S^2 = 0$. In what follows $u_S = \frac{q}{c}\eta_S^2$ and $\alpha + \beta^*\eta_S^2 + \gamma\eta_S^4 = 0$, where $\beta^* = \left(\beta - 2\frac{q^2}{c}\right)$. So that $\eta_S^2 = \frac{1}{2\gamma}\left(\sqrt{\beta^{*2} - 4\alpha\gamma} - \beta^*\right)$ for the first order phase transitions (where $\beta^* < 0, \gamma > 0$) and $\eta_S^2 = -\alpha/\beta^*$ for the second order phase transitions (where $\beta^* > 0, \gamma = 0$). Note that the condition $c\beta > 2q^2$ should be valid for the second order phase transitions. The solution of linearized equations (5) can be found after elementary transformations in the following form:

$$\tilde{\eta} = \tilde{\chi}(k)\tilde{E}, \qquad \tilde{u} = -\frac{(ifk - 2q\eta_S)}{(c + vk^2)}\tilde{\chi}(k)\tilde{E} \qquad (6)$$

The generalized susceptibility (Green function) is introduced here:

$$\tilde{\chi}(k) = \left(\alpha + 3\beta\eta_S^2 + 5\gamma\eta_S^4 - 2qu_S + gk^2 + wk^4 - \frac{4q^2\eta_S^2 + f^2k^2}{c + vk^2}\right)^{-1}. \qquad (7)$$

The condition of the spatial modulation appearance with a period $k$ is:

$$\alpha + 3\beta\eta_S^2 + 5\gamma\eta_S^4 - 2qu_S + gk^2 + wk^4 - \frac{4q^2\eta_S^2 + f^2k^2}{c + vk^2} = 0 \qquad (8)$$

### 4.1. Flexocoupling induced SMPs in commensurate ferroics

In a high temperature parent phase of **commensurate ferroics, for which g>0 and w=0**, the values $\alpha > 0$, $\eta_S = 0$, $u_S = 0$. For the case the condition (8) transforms into the equivalent equation $(\alpha + gk^2)(c + vk^2) - f^2k^2 = 0$. Introducing the dimensionless variables,

$$k^* = \sqrt{\frac{2v}{c}}k, \qquad f^* = \frac{f}{\sqrt{cg}}, \qquad v^* = \frac{\alpha v}{cg}, \qquad (9)$$

the solution of Eq.(8) is

$$k_{\pm}^{*}=\sqrt{f^{*2}-1-v^{*}\pm\sqrt{\left(f^{*2}-1-v^{*}\right)^{2}-4v^{*}}}\,. \tag{10}$$

Under the condition $f^{*2}\geq 1+v^{*}$ the SMP with a minimal and maximal modulation periods, $k_{+}^{*}$ and $k_{-}^{*}$, appears. Under the condition $f^{*2}<1+v^{*}$ the static homogeneous state is thermodynamically stable in the parent phase.

In a **low temperature long-range ordered phase of commensurate ferroics**, where $\alpha<0$, $\eta_S\neq 0$, $u_S\neq 0$, introducing the renormalized positive parameter $\alpha_S=\alpha+\left(3\beta-2\dfrac{q^2}{c}\right)\eta_S^2+5\gamma\eta_S^4$ in Eq.(8), it reads as $\left(\alpha_S+gk^2\right)\left(c+vk^2\right)-f^2k^2-4q^2\eta_S^2=0$. Using the dimensionless variables similar to those in Eq.(9)

$$k^{*}=\sqrt{\dfrac{2v}{c}}k,\quad f^{*}=\dfrac{f}{\sqrt{cg}},\quad v_S^{*}=\dfrac{\alpha_S v}{cg},\quad Q_S^2=4\dfrac{q^2\eta_S^2}{\alpha_S c}, \tag{11}$$

we get:

$$k_{\pm}^{*}=\sqrt{f^{*2}-1-v_S^{*}\pm\sqrt{\left(f^{*2}-1-v_S^{*}\right)^{2}-4v_S^{*}\left(1-Q_S^2\right)}}\,. \tag{12}$$

Under the condition $f^{*2}<1+v_S^{*}$ the static homogeneous state is thermodynamically stable in the parent phase. Under the condition $f^{*2}\geq 1+v_S^{*}$ the SMP with a minimal and maximal modulation periods given by expression (12) appears.

**4.2. Flexocoupling impact on SMPs periods in incommensurate ferroics**

`In a high temperature parent phase of **incommensurate ferroics, for which g<0 and w>0,** the values $\alpha>0$, $\eta_S=0$, $u_S=0$, the condition (8) transforms into the equivalent equation is $\left(\alpha+gk^2+wk^4\right)\left(c+vk^2\right)-f^2k^2=0$. Introducing the dimensionless variables:

$$k^{*}=\sqrt{\dfrac{2(gv+cw)}{-cg}}k,\quad f^{*}=\dfrac{f}{\sqrt{-cg}},\quad v^{*}=-\dfrac{\alpha v}{cg},\quad w^{*}=-\dfrac{cw}{vg} \tag{13}$$

we get the solution:

$$k_{\pm}^{*}=\sqrt{1+f^{*2}-v^{*}\pm\sqrt{\left(1+f^{*2}-v^{*}\right)^{2}-4v^{*}\left(w^{*}-1\right)}}\,. \tag{14}$$

Assuming that $(gv+cw)>0$, the static homogeneous state is thermodynamically stable in the parent phase under the condition $f^{*2}<v^{*}-1$. Under the condition $f^{*2}\geq v^{*}-1$ the SMP phase with a minimal and maximal modulation periods, $k_{+}^{*}$ and $k_{-}^{*}$, given by Eq.(14) appears.

In a **low temperature long-range ordered phase of incommensurate ferroics**, where $\alpha<0$, $\eta_S\neq 0$, $u_S\neq 0$, Eqs.(8) reads $\left(\alpha_S+gk^2+wk^4\right)\left(c+vk^2\right)-f^2k^2-4q^2\eta_S^2=0$. Introducing positive dimensionless variables:

$$k^* = \sqrt{\frac{2(gv+cw)}{-cg}}k, \quad f^* = \frac{f}{\sqrt{-cg}}, \quad v_S^* = -\frac{\alpha_S v}{cg}, \quad w^* = -\frac{cw}{vg}, \quad Q_S^2 = 4\frac{q^2\eta_S^2}{\alpha_S c} \quad (15)$$

we get:

$$k_\pm^* = \sqrt{1+f^{*2}-v_S^* \pm \sqrt{(1+f^{*2}-v_S^*)^2 - 4v_S^*(w^*-1)(1-Q_S^2)}}. \quad (16)$$

Under the condition $f^{*2} < v_S^* - 1$ the static homogeneous state is thermodynamically stable in the parent phase. Under the condition $f^{*2} \geq v_S^* - 1$ the SMP phase appears.

Reference values of LDG-expansion coefficients and other parameters introduced in Eqs.(9), (11), (13) have been estimated for ferroics with incommensurate and commensurate phases in Ref.[57]. Namely the parameters $\alpha_S=(0-2)\times 10^3$ C$^{-2}$·mJ, $\beta$ changes from $-5\times 10^8$ J C$^{-4}$·m$^5$ for the ferroics with the first order phase transitions to $+1\times 10^8$ J C$^{-4}$·m$^5$ for the ferroics with the second order phase transitions, $\gamma$ changes from $10^9$ J C$^{-6}$·m$^9$ to $10^{11}$ J C$^{-6}$·m$^9$, $\eta_S$ changes from 0 to 0.7 C/m$^2$ in dependence on temperature, $q$ is about $10^9$ Vm/C, $c$ is about $(1-10)\times 10^{10}$ Pa. The gradient coefficient $g$ changes from $-6\times 10^{-10}$C$^{-2}$m$^3$J for ferroics with incommensurate phase like Sn$_2$P$_2$Se$_6$ to $+5\times 10^{-10}$C$^{-2}$m$^3$J for commensurate ferroics like Sn$_2$P$_2$S$_6$. The higher gradient $w=0$ for commensurate ferroics and is about $(1-3)\times 10^{-27}$ J·m$^5$/C$^2$ for ferroics with incommensurate phase. Flexocoefficient $f$ changes from $-5$ V to $+5$ V, $v$ is within the range $(10^{-7}-10^{-6})$V s$^2$/m$^2$. Using the reference values we estimated that dimensionless parameters changes in the range $Q^* = (0.02 - 0.8)$, $w^* = 0$ for commensurate ferroics and $w^* = -(2-20)$ for ferroics with incommensurate phase, $f^* = (-5 - +5)$, $v^* = (-5 - +5)$ and $k^* = (0-10)$. Thus for a reasonable range of striction, nonlinearity and elastic stiffness coefficients the inequality $0 \leq Q^* < 1$ is likely to be valid.

### 4.3. Discussion of analytical results

Diagram of the stable phases, which exist in commensurate and incommensurate ferroics with SMP is shown in **Figure 2,** and the phase boundary separating SMP and HP phases is shown by solid curvefor commensurate ferroics (with $g>0$ and $w=0$) and dashed curve for incommensurate ones (with $g<0$ and $w>0$). As one can see, the minimal absolute value of the flexoconstant $|f_{cr}^*| = 1$ is required for SMP appearance in commensurate ferroics. That say the HP is absolutely stable in commensurate ferroics at $|f^*|<1$ and $v^* = 0$, at that the value $|f_{cr}^*|$ monotonically increases with $v^*$ increasing (see Ref.[57] and red solid curve in **Fig.2**). In contrast to commensurate ferroics, the SMP appears at $f^* = 0$ and $v^* \geq 1$ in the incommensurate ferroics. Then the maximal absolute value $f_{cr}^*(v^*)$ appears and its absolute value increases quasi-linearly with $v^*$ increasing (see Ref.[57] and blue dashed curve in **Fig.2**).

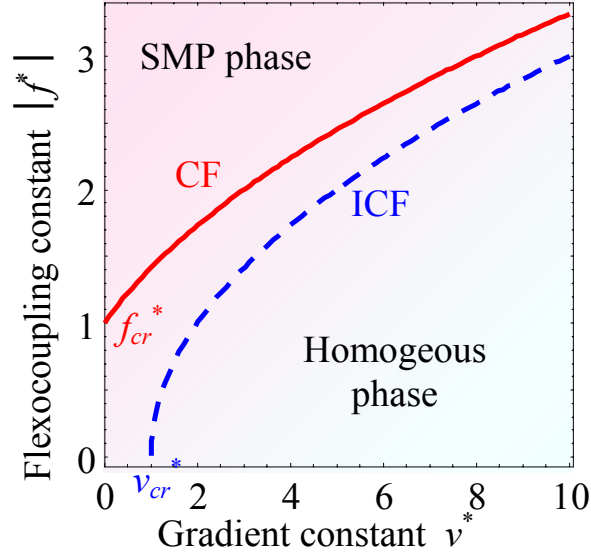

**Figure 2.** Diagram of the SMP phase stability in commensurate (solid curve $|f^*| = \sqrt{v^*+1}$ with legend CF) and incommensurate (dashed curve $|f^*| = \sqrt{v^*-1}$ with legend ICF) ferroics plotted in coordinates $f^*$ and $v^*$.

Dependences of the wave vectors $k_+^*$ and $k_-^*$ on the dimensionless flexoconstant $f^*$ and gradient constant $v^*$ calculated for *commensurate* ferroics are shown in **Figs. 3.** The curves behavior is very similar in the parent ($\alpha > 0$) and long-range ordered ($\alpha < 0$) phases [e.g. compare plots (a) and (b), or (c) and (d), correspondingly]. A "gap" (i.e. the absence of $k_\pm^*$ at $f^* < f_{cr}^*$) exists for all curves corresponding to different $v^*$-values [see **Figs.3(a)** and **3(b)**]. The gap width $d$ is conditioned by the value $f_{cr}^*(v^*)$ and it increases with $v^*$ increasing [compare different curves in **Figs.3(a)** and **3(b)**]. At $f^* > f_{cr}^*$ the wave vector $k_-^*$ decreases and $k_+^*$ increases with $f^*$ increasing [compare dashed and solid curves in **Figs.3(a)** and **3(b)**]. There is the maximal positive value $v_{cr}^*(f^*)$ of SMP appearance at fixed $f^*$ value [see **Figs.3(c)** and **3(d)**]. At $v^* < v_{cr}^*$ wave vector $k_-^*$ increases and $k_+^*$ decreases with $v^*$ increasing [compare dashed and solid curves in **Figs.3(c)** and **3(d)**].

Dependences of the wave vectors $k_+^*$ and $k_-^*$ on the dimensionless flexoconstant $f^*$ and gradient constant $v^*$ calculated for *incommensurate* ferroics are shown in **Figs.4.** A gap $d$ exists only for the two curves corresponding to the highest $v^*$-values [see in **Figs 4(a)** and **4(b)**], as it follows from the dependence $f_{cr}^*(v^*)$ analyzed in Refs.[57-58]. The wave vector $k_-^*$ decreases with $|f^*|$ increasing, while $k_+^*$ increases with $|f^*|$ increasing [compare dashed and solid curves in **Figs 4(a)** and **3(b)**]. **Figs.4(c)** and **4(d)** plotted for incommensurate ferroics show the same behavior as the **Figs.3(c)** and **3(d)** for commensurate ferroics. Actually, there is the *maximal positive* value $v_{cr}^*(f^*)$ of SMP appearance at fixed $f^*$ value [see **Figs.4(c)** and **4(d)**]. At $v^* < v_{cr}^*$ wave vector $k_-^*$ increases and $k_+^*$ decreases with $v^*$ increasing [compare dashed and solid curves in **Figs.4(c)** and **4(d)**].

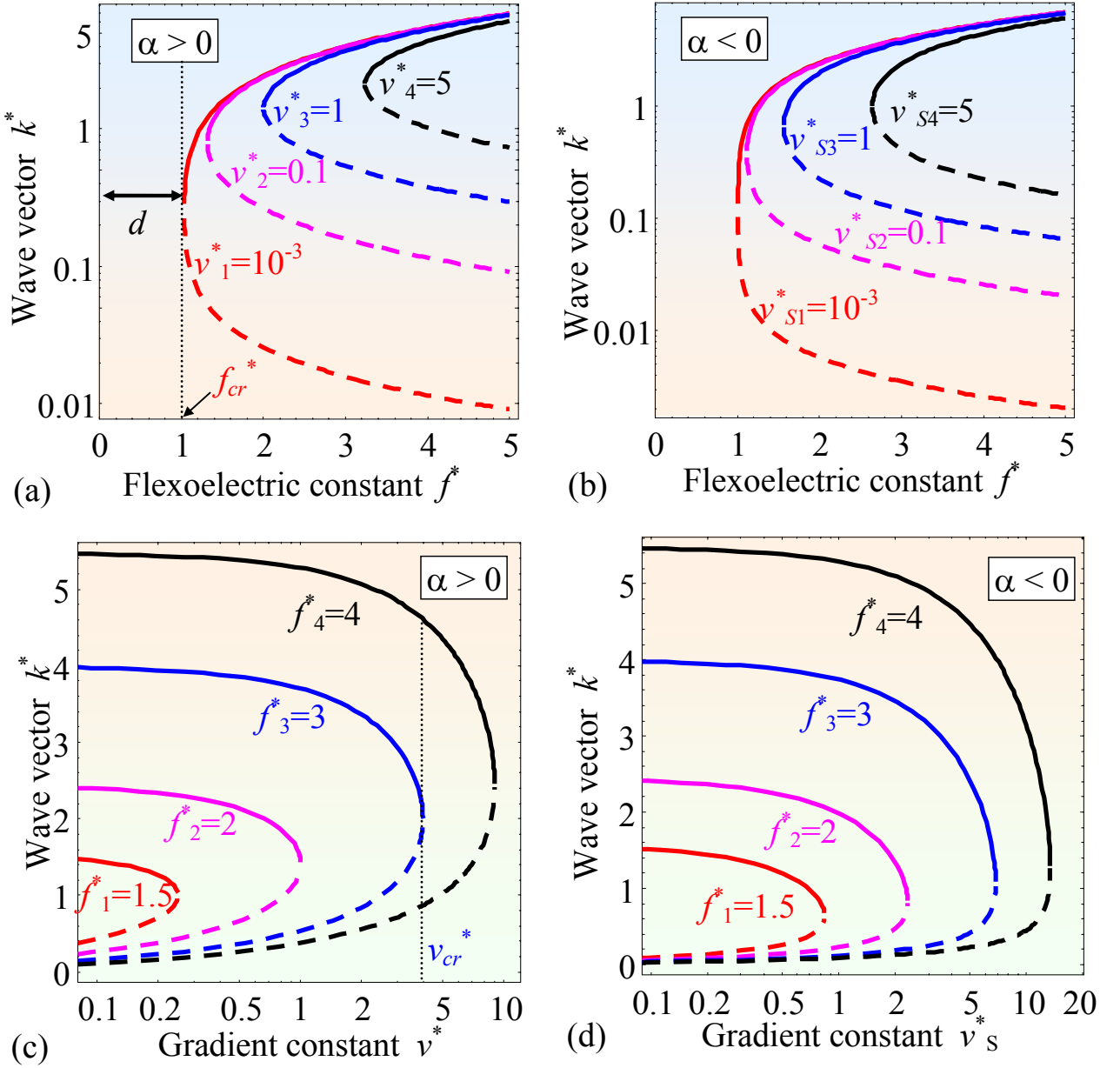

**Figure 3. Commensurate ferroics.** Dependences of the wave vectors $k_+^*$ (solid curves) and $k_-^*$ (dashed curves) on the flexoelectric constant $f^*$ **(a,b)** and elastic strain gradient constant $v^*$ **(c)** and $v_S^*$ **(d)**. Different plots are calculated for $\alpha > 0$ **(a,c)** and $\alpha < 0$ **(b,d)**. Different curves in each of the plots are calculated for several values of $f_i^*$, $v_i^*$ and $v_{Si}^*$ listed in the legends at the plots. Parameter $Q_S^2 = 0.95$ for the plots **(b)** and **(d)**.

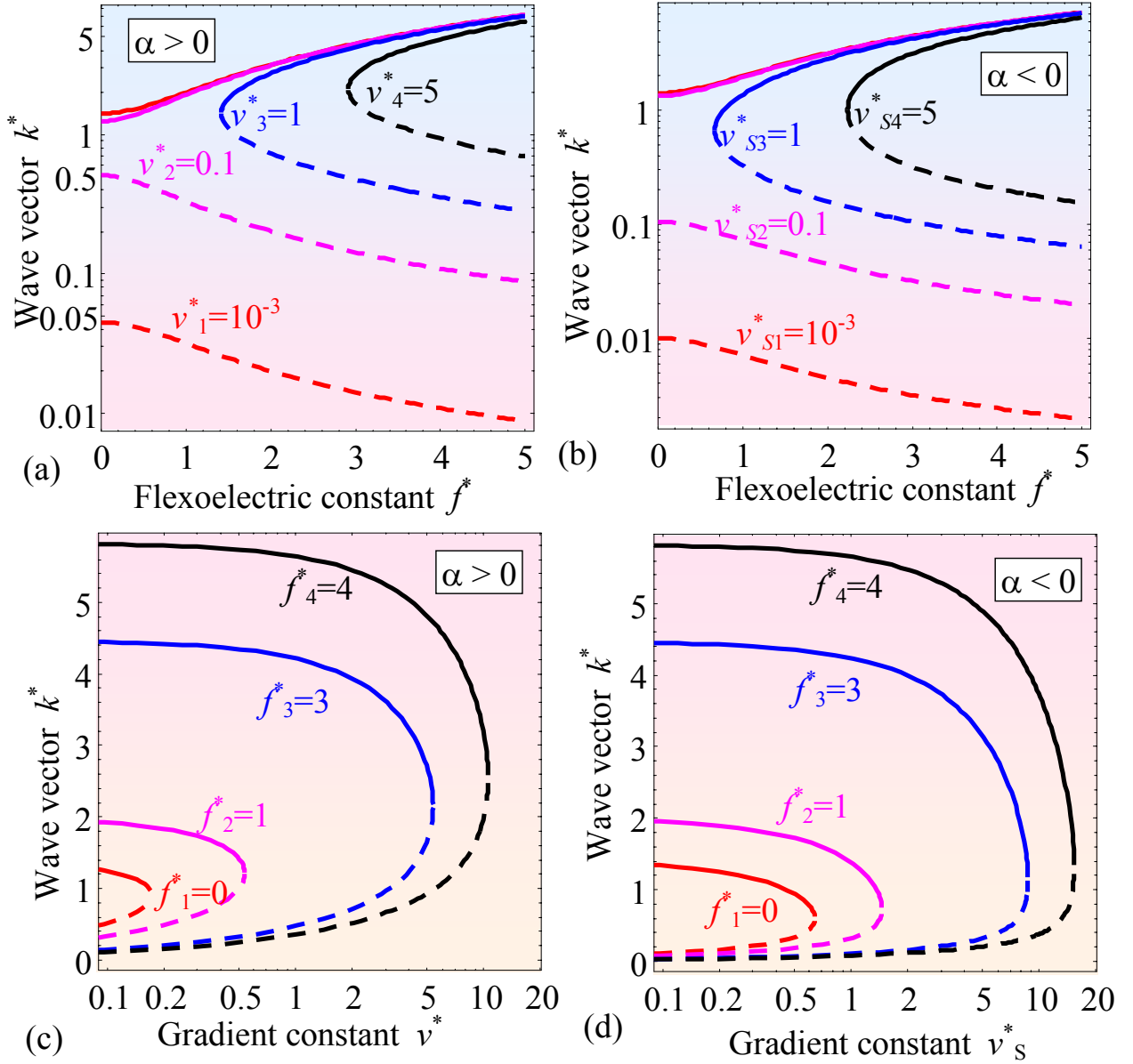

**Figure 4. Incommensurate ferroics.** Dependences of the wave vectors $k_+^*$ (solid curves) and $k_-^*$ (dashed curves) on the flexoelectric constant $f^*$ **(a,b)** and elastic strain gradient constant $v^*$ **(c)** and $v_S^*$ **(d)**. Different plots are calculated for $\alpha > 0$ **(a,c)** and $\alpha < 0$ **(b,d)**. Different curves in each of the plots are calculated for several values of $f_i^*$, $v_i^*$ and $v_{Si}^*$ listed in the legends at the plots. Parameters $w^* = 2$ and $Q_S^2 = 0.95$ for the plots **(b)** and **(d)**.

## 5. Flexoelectricity in liquid crystals

Both theoretical and experimental study of the flexoelectricity in liquid crystals (**LC**) is a challenge [61, 62, 63], at that corresponding research approaches should be adapted for three main types of liquid crystals: nematic, smectic and cholesteric [64]. Among them the flexoeffect impact is relatively well-studied in nematics [65, 66].

In a nematic liquid crystal (**NLC**) molecular long axes are oriented approximately parallel to each other, but the centers of mass of the molecules are located randomly, as in the liquid. The direction of dominant orientation is characterized by the order parameter - director vector **n**, which bistable states, **n** and **–n,** are energetically equivalent in the absence of external electric field. NLC, even if they consist of polar molecules, are non-polar substances, which can be polarized by an electric field **E** [67]. The interaction with the field is analogous to the piezoelectric effect in solids, which consists in the appearance of polarization in the creation of elastic stresses.

On the other hand, electric polarization is induced by deformations of the longitudinal and transverse bends. The effect was firstly predicted theoretically by Meyer in 1969 [68] and then revealed experimentally. This direct coupling is a direct flexoelectric effect in LC. Similarly to solids, there is also a converse static flexoeffect in soft matter, when the applied external field polarizing the NLC, creates distortions in the director distribution. That say the static flexoelectric effect in the NLC is the coupling of the electric field with the spatial gradient of the director field, $\vec{P}^{flexo} \sim f \cdot \vec{n}(\nabla \vec{n})$, where $f$ is the flexoelectric coefficient that couples the polarization vector with the longitudinal or transverse bending. Corresponding contribution of flexoelectricity to the thermodynamic potential is proportional to the strength of the applied electric field (similar to the piezoelectric effect in solids). The information about dynamic flexoeffect in LC is absent.

In contrast to the well-studied Fredericks effect, caused by the anisotropy of the dielectric constant and being quadratic to the applied electric field, the flexoelectric effect linear in the electric field caused in NLC was revealed in 1969 [67]. It is conditioned by the by the distortion of the director field in NLC. Similarly to the omnipresence of flexoelectric effect in solids, it turned out that the flexoelectric polarization is a universal property of NLC [61]. The physical picture of the flexoelectric effect in NLC is shown in **Figs. 5(a)** and **(b),** respectively. Polar molecules change their orientation in external electric filed or due to elastic stimuli and their net dipole moments appears.

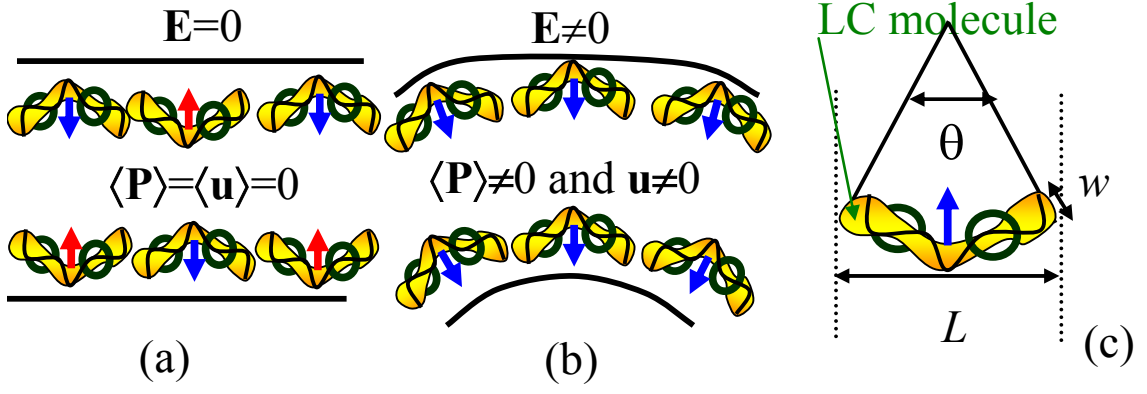

**Figure 5.** Illustration of the nature of the flexoelectric effect for the bend-shaped LC molecules. Red and blue arrows indicate the dipole moments of the bended polar LC molecules. **(a)** External electric field or strain are absent (**E**=0 and **u**=0). **(b)** External electric field (or strain) is present (**E**≠0 and **u**≠0).

If the dielectric anisotropy of NLC, $\varepsilon_a K$ is small enough (satisfying the conditions listed in Ref.[61]) then e.g. the external electric field $\mathbf{E}_z$ changes the orientation of director $\mathbf{n}_x$ in the cell for the fields higher than the threshold value, $E > E_t$ where $E_t$ is the threshold value of the field. $E_t$ is inversely proportional to the flexoelectric coefficient *f* and given by expression (see Ref.[61]):

$$E_t = \frac{2\pi K}{|f|d}\left(1 - \frac{\varepsilon_a K}{4\pi f^2}\right)^{-1} \quad (17)$$

The size of LC cell is *d*. From the formula (17) the threshold value of the field is the smaller the larger is $|f|$. Furthermore a planar oriented LC with rigid boundary conditions at $z = \pm d/2$ in an external field, which exceeds the threshold value $E > E_t$, forms a periodic structure due to the flexoelectric effect. If the dielectric anisotropy is not small, then instead of the periodic structure, the usual Fredericks transition takes place [61]. Note that relatively small changes in molecular shapes can lead to strong changes in their polar properties. For instance the first observations found ferroelectric polarization switching [69, 70] in bend-shaped LC molecules, but later it appeared that the switching reveals antiferroelectric behavior [71, 72, 73, 74, 75, 76, 77, 78, 79].

**5.2. Estimates of flexoelectric polarization and coefficients in LCs**

Flexoelectricity as defined by Meyer [68] for NLC with bend-shaped e molecules is a special case of the general definition Eq.(1), where the gradient of the strain is replaced by the gradient of the director **n**. It was assumed that the constituent molecules of the NLCs are rotate freely around their axes; and so their average dipole moments is zero in the absence of electric fields, and the net polarization of the material is also zero [61]. When LC consists of polar bend-shaped molecules they can become macroscopically polar. It follows from symmetry considerations that the strain-induced flexoelectric polarization $\vec{P}^{flexo}$ can be written as [68, 61]:

$$\vec{P}^{flexo} = f_1\vec{n}(\nabla\vec{n}) + f_3\vec{n}(\nabla\times\vec{n})\times\vec{n} \tag{18}$$

The value of polarizaion $\vec{P}^{flexo}$ is characterized by two flexoelectric coefficients, $f_1$ and $f_3$, for splay and bend, respectively. The microscopic origin of these phenomenological flexocoelectric cefficients contains at least two different mechanisms, dipolar and quadrupolar, exist (see details in Refs.[61] and [80]).

The estimates below are based on the results of Ref.[61].The dipolar mechanism is sensitive to the molecular shape. By dimensional considerations one can estimate the flexocoefficients due to dipolar mechanism as $|f_{1,3}| \leq \mu_e/a^2$, where $\mu_e \sim (3-15)\times10^{-30}$ C·m is the molecular dipole moment and $a \sim (2-4)$ nm. This means that $f_1$ and $f_3$ are expected to be of the order of pC/m. Hence these estimates show that the dipolar flexoelectric coefficients [being of the order of (1–10) pC/m] are too small for practical applications. However the estimates give essentially underestimated values, as discussed below.

Assuming a random 3D-distribution of the centre of masses of the bent-core NLC molecules, Helfrich [81, 82], Derzhanski and Petrov [83] derived another (more rigorous) expression for the observable bending flexoelectric coefficient [61], $f_3 = \frac{\mu_\perp K_3}{2k_BT}\theta\sqrt[3]{N\left(\frac{w}{L}\right)^2}$. Here $\mu_\perp$ is the dipole moment perpendicular to the long axis, $K_3$ is the bend elastic constant, $\theta$ is the bending angle (see **Fig. 5(c)** and Ref.[61]), $N$ is the concentration number, $L$ is the length, $w$ is the width of the molecules, $k_BT$ is the thermal energy. The measurements of $f_3$ in calamitic NLCs revealed essentially higher values than expected from the estimate, indicating that not only the dipolar mechanism contributes to flexoelectricity, and it is also caused by the interaction between the electric field gradient and the molecular quadrupole moments [84, 85]. Molecules of band-core NLCs are characterized by an angle $\theta_0 \approx 60$ degree, which is much larger than $\theta_0$ of calamitic NLCs [61]. Therefore in these compounds the dipolar contribution to $f_3$ can be higher at least in an order of magnitude, while $f_1$ may be of the same order as estimated. Moreover, band-core nematics exhibit a giant flexoelectric effect with $f_1$ much larger than expected from the above estimates [61].

### 5.3. Free energy functional with flexoelectric coupling in liquid crystals and SMPs

Using the definition of the flexoelectric coupling, Eq.(18), introduced in the subsection 5.2, the bulk part of the total free energy could be decomposed in several parts, namely bulk part, determining symmetry of ordered phase and its dependence on temperature *T*; distortion energy and electrostatic/flexoelectric contributions:

$$G_V = \int d^3r \left( g_{bulk}(T,S,\vec{n}) + g_{elastic}(S,\vec{n}, \vec{\nabla} \otimes \vec{n}) + g_{flexoelect}(S,\vec{n}, \vec{\nabla} \otimes \vec{n}) \right) \quad (19)$$

Here we introduced scalar degree of ordering *S* and unit vector $\vec{n}$ determining local direction of symmetry axis (vector field of director). Distortion energy contribution ("elastic" energy) could be expressed via director vector $\vec{n}$ and its spatial derivatives as follows [67]:

$$g_{elastic} = \frac{K_1}{2}(div(\vec{n}))^2 + \frac{K_2}{2}(\vec{n} \cdot rot(\vec{n}))^2 + \frac{K_3}{2}(\vec{n} \times rot(\vec{n}))^2 \quad (20)$$

Here $K_1$, $K_2$ and $K_3$ are elastic coefficients corresponding to transversal bending, twist deformation of director field and longitudinal bending respectively, these parameters are proportional to the square of ordering degree *S*. Electrostatic field $\vec{E}$ contribution to the free energy at the fixed potentials of electrodes is (see e.g. [67]):

$$\begin{aligned} g_{flexoelect} &= -f_1(\vec{E} \cdot \vec{n})div(\vec{n}) - f_3(\vec{E} \cdot (\vec{n} \times rot(\vec{n}))) - \frac{\varepsilon_0 \varepsilon_b}{2}(\vec{E})^2 \\ &\equiv -f_1(\vec{E} \cdot \vec{n})div(\vec{n}) + f_3(\vec{E} \cdot (\vec{n} \cdot \vec{\nabla}) \cdot \vec{n}) - \frac{\varepsilon_0 \varepsilon_b}{2}(\vec{E})^2 \end{aligned} \quad (21)$$

Here we introduced $f_1$ and $f_3$ are flexoelectric coefficients corresponding to longitudinal and transversal bending. In the latter transformation we take into account that $\vec{n} \cdot \vec{n} \equiv 1$.

The surface energy contribution has the form:

$$G_S = \int_S d^2r \left( \frac{K_1}{2h_1} n_1^2 + \frac{K_2}{2h_2} n_2^2 + \frac{K_3}{2h_3} n_3^2 \right), \quad (22)$$

where the analog of the extrapolation lengths $h_i$ are introduced (*i*=1, 2, 3).

Euler-Lagrange equation for the director vector is:

$$\frac{\partial (g_{bulk} + g_{elastic} + g_{flexoelect})}{\partial n_i} - \frac{\partial}{\partial x_j} \left( \frac{\partial (g_{elastic} + g_{flexoelect})}{\partial n_{i,j}} \right) = \lambda n_i \quad (23)$$

Here we introduced Lagrange multiplier λ to take into account the condition $\vec{n} \cdot \vec{n} \equiv 1$ and used short hand designation for tensor of gradient of director field, $n_{i,j} \equiv \partial n_i / \partial x_j$. Evident form of (S1.4) is rather cumbersome (see e.g. [67]). Only for the case $K_1 = K_2 = K_3$ one could get simple expression:

$$\frac{\partial (g_{elastic})}{\partial \vec{n}} - \frac{\partial}{\partial x_j} \left( \frac{\partial (g_{elastic})}{\partial \vec{n}_{,j}} \right) = -K \Delta \vec{n} \quad (24)$$

For instance, flexoelectric term contribution is

$$\frac{\partial g_{flexoelect}}{\partial \vec{n}} - \frac{\partial}{\partial x_j}\left( \frac{\partial g_{flexoelect}}{\partial \vec{n}_j} \right) = -f_1\left[ \vec{E} \, div(\vec{n}) - \vec{\nabla}(\vec{n} \cdot \vec{E}) \right] + f_3\left[ (\vec{E} \cdot (\vec{\nabla} \otimes \vec{n})) - \vec{E} \, div(\vec{n}) - (\vec{E} \cdot \vec{\nabla})\vec{n} \right] \quad (25)$$

Using the formalism Eqs.(19)-(25) with the natural boundary conditions for polarization and director, at the LC film boundaries $z = \pm d/2$, namely

$$\left.\frac{\partial \theta}{\partial z} - \frac{1}{h_3}\sin\theta\cos\theta = 0\right|_{z=-d/2}, \quad \left.\frac{\partial \theta}{\partial z} + \frac{1}{h_3}\sin\theta\cos\theta = 0\right|_{z=+d/2}, \quad (26)$$

we calculated and analyzed the influence of the flexoelectric coupling on the director orientation and induced polarization in a NLC film. Note that $n_1 = \cos\theta\cos\varphi$, $n_2 = \cos\theta\sin\varphi$ and $n_3 = \sin\theta$, Polarization is given by Eq.(18), namely, $\vec{P}^{flexo} = f_1\vec{n}(\nabla\vec{n}) + f_3\vec{n}(\nabla\times\vec{n})\times\vec{n}$.

Profiles of the director angle θ and polarization calculated for a 20-nm film are shown in **Fig.6(a)** and **6(b)**, respectively. It is seen from the figures that without flexoelectric coupling (*f*=0) any oscillations or maxima, or any other features are absent at the director and polarization distributions [see black curves in **Figs.6**]. Relatively small flexoelectric coupling with the strength ($0.3\text{V} < f < 0.5\text{V}$) leads to the small maximums on the director angle θ profiles near the LC boundaries [see blue and magenta curves in **Figs.6(a)**]. At the same time the features (increase or maxima) on the profiles of flexo-polarization are noticeable [see blue and magenta curves in **Figs.5(b)**]. Flexoelectric coefficient $f \geq 0.7\text{ V}$ induces the LC film transition into a spatially modulated phase with quasi-harmonic modulation of the director angle $\theta \cong \theta_0 \cos(qz)$ and polarization $P \cong P_0 \cos(qz)$ [see red curves in **Figs.6**].

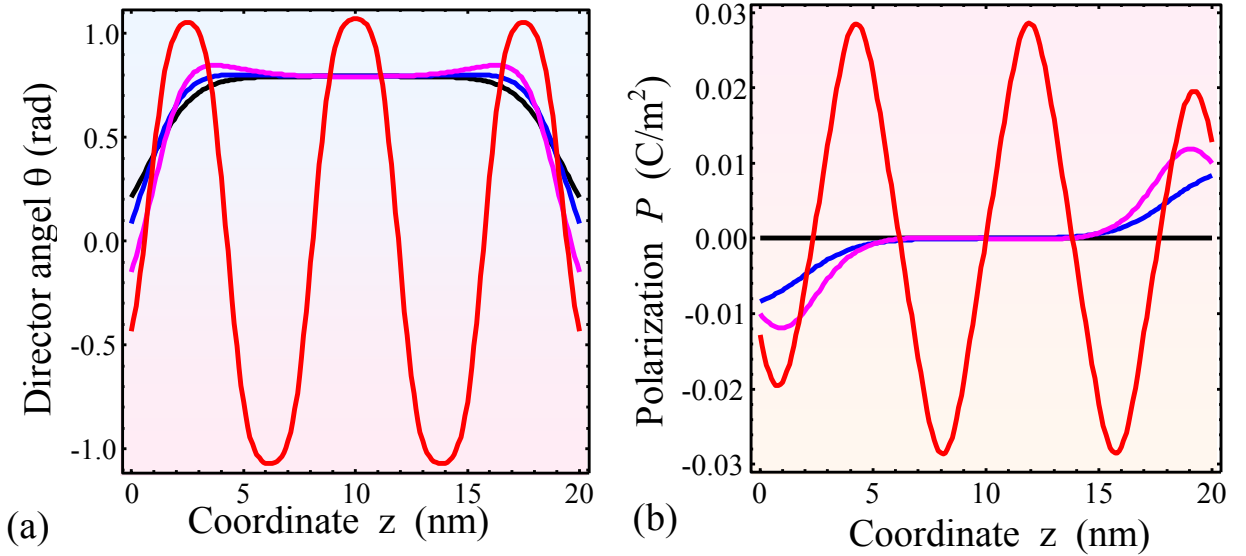

**Figure 6. Influence of the flexoelectric coupling on the director orientation and polarization in a NLC.** Order parameter – director angle θ **(a)** and polarization **(b)** distribution across the LC film of thickness 20 nm calculated for different values of flexoelectric coefficient *f* = 0, 0.3, 0.50, 0.7 V (black, blue, magenta, red curves respectively).

Therefore an oriented LC film with the boundary conditions (25) imposed at the cell boundaries forms a spatially-modulated periodic structure due to the flexoelectric effect without any external field. If the flexocoupling strength is relatively small in comparison dielectric anisotropy energy, the usual Fredericks transition takes place in external electric field instead of the spatially-modulated structure.

Note that the spatially modulated polarization phases have been firstly revealed in a smectic liquid crystal [61, 86]. As discussed by Bailey and J´akli, [87] a steric or electrostatic inclusion in the core of NLC molecules leads to layer modulation [61], in which the broken smectic ribbons are separated by melted fluid nematic regions (see Ref.[61] and refs [88, 89, 90] therein on the small-angle X-ray scattering (SAXS) measurements). These smectic clusters are at least partially ferroelectric and they can exhibit the soft mode-type low-frequency dielectric relaxations [91] and the giant flexoelectric effects [92, 93].

## 6. Conclusions

In the review we briefly analyze the state-of-art in the theory of flexoelectric phenomena and analyze how significantly the flexoelectric coupling can change the polar order parameter distribution in ferroics and liquid crystals. The special attention in paid to the appearance of the spatially modulated phases induced by the flexocoupling in solids and soft matter. Results of theoretical modeling performed in the framework of the Landau-Ginzburg-Devonshire formalism revealed that the general feature is the appearance of the spatially modulated phases in ferroics and liquid crystals taking place with increasing of the flexocoupling strength. We'd like to underline that theoretical and experimental study of flexoelectricity and related phenomena in nanosized and bulk ferroics, liquid crystals are very important for their advanced applications in nanoelectronics, memory devices and LC displays.

**Acknowledgments.** M.V.S. and D.V.K acknowledge MK-1720.2017.8, BRFFR (# F16R-066) RFBR (# 16-58-00082).